\documentstyle[12pt] {article}
\newcommand{\beq}{\begin{equation}}
\newcommand{\eeq}{\end{equation}}
\newcommand{\beqa}{\begin{eqnarray}}
\newcommand{\eeqa}{\end{eqnarray}}
\newcommand{\ba}{\begin{array}}
\newcommand{\ea}{\end{array}}

\begin{document}

\begin{center}
{\large \bf Instability and Chaos \\
in Spatially Homogeneous Field Theories} \\
\end{center}

\vskip 1. truecm

\begin{center}
{\bf Luca Salasnich}\footnote{Electronic address: 
salasnich@mi.infm.it}
\vskip 0.5 truecm
Istituto Nazionale per la Fisica della Materia, Unit\`a di Milano, \\
Dipartimento di Fisica, Universit\`a di Milano, \\
Via Celoria 16, 20133 Milano, Italy \\
\end{center}

\vskip 1.5 truecm

\begin{center}
{\bf Abstract}
\end{center}
\par
Spatially homogeneous field theories are studied in the framework of 
dynamical system theory. In particular we consider 
a model of inflationary cosmology and a Yang--Mills--Higgs system. 
We discuss also the role of quantum chaos and 
its application to field theories. 

\vskip 1.5 truecm
\par

\begin{center}
PACS Numbers: 11.15.-q; 98.80.Cq; 05.45.+b
\end{center}

\newpage

\section{Introduction}

\par
Quantum field theory offers a wide variety of applications, in particular  
for condensed matter$^{1}$ and elementary particle physics$^{2}$. 
Field theoretic ideas also reach for the cosmos through the 
development of the inflationary scenario -- a speculative, 
but completely physical analysis of the early universe, 
which appears to be consistent with available observations$^{3}$.
\par
In the last years there has been much interest in the chaotic behaviour 
of field theories$^{4-8}$. In this paper we discuss and extend 
our recent results$^{12-17}$ on instability and chaos 
in classical and quantum field theory. 
In Section 2 we show how spatially homogeneous field theories can be studied 
by using the dynamical system theory and we 
introduce some basic definitions for the regular and chaotic dynamics 
of classical and quantum systems. In Section 3 we analyze 
the local stability of a inflationary scalar field minimally 
coupled to gravity and its point attractors in the phase space. 
The value of the scalar field in the vacuum is a bifurcation parameter 
and we discuss the existence of a stable limit cycle. 
Finally, in Section 4 we study the spatially homogenous 
SU(2) Yang--Mills--Higgs system. We show that for this system 
a classical order--chaos transition occurs both in classical and 
quantum mechanics. 

\section{Field theories as dynamical systems}

\par
In this Section we introduce some basic ideas of the dynamical system theory. 
We clarify the concept of ergodic system giving a hierarchy of chaos. 
\par
Let us consider a classical relativistic 
scalar field theory with action
\beq
S[\phi ] = \int d^4x \; L(\phi , \partial_{\mu} \phi ) \; ,
\eeq
where $L$ is the Lagrangian density of the system, 
$\partial_{\mu}=({\partial\over \partial t},{\bf \nabla})$ is the 
covariant derivative and $\phi = \phi (x)$ is a real scalar 
field with $x_{\mu}=(t,{\bf x})$ the space--time position$^{2}$. 
It is well known that by imposing the Hamilton's Least Action 
Principle 
\beq
\delta S[\phi ] = 0 \; ,
\eeq 
we obtain the Euler--Lagrange equation of motion of the system
\beq
{\partial L\over \partial \phi}-\partial_{\mu}{\partial L\over 
\partial (\partial_{\mu} \phi )} = 0 \; .
\eeq
The homogenous space approximation means that we can neglect the 
spatial dependence of the field, thus we can perform the following 
substitution: 
\beq
\phi (t,{\bf x}) \to \phi (t) \; ,
\eeq
and the resulting equation of motion is given by 
\beq
{\partial L\over \partial \phi}-{d\over dt}{\partial L\over 
\partial {\dot \phi} } = 0 \; .
\eeq
By introducing the momentum 
\beq
\chi = {\partial L\over \partial {\dot \phi} } \; ,
\eeq
and the Hamiltonian
\beq
H (\chi , \phi )= {\dot \phi} \chi - L(\phi , {\dot \phi }) \; , 
\eeq
the second order equation of motion can be 
written as a system of two first order Hamilton's equations
\beq
{\dot {\bf z}}={\bf f}({\bf z})
\eeq
where ${\bf z}=(\phi , \chi )$ is a point in a two dimensional 
phase space and ${\bf f}=(f_1,f_2)$ is given by:
\beq
f_1(\phi ,\chi )= {\partial H\over \partial \phi} \; ,
\quad 
f_2(\phi ,\chi )= -{\partial H\over \partial \chi} \; .
\eeq
\par
This is a general result: any homogenous field theory can be 
written as a system of $N$ first order differential equations, i.e. 
a dynamical system. In the next Sections we shall consider 
non--conservative and non--Abelian field theories. 

\subsection{Dynamical System Theory} 

\par 
A dynamical system is defined by N first order differential equations 
\beq
{\dot {\bf z}}(t)={\bf f} ({\bf z}(t),t) \; , 
\eeq
where the variables ${\bf z}=(z_{1},...,z_{N})$ are in the phase
space $\Omega$ (the euclidean space $R^{N}$, unless otherwise 
specified). These equations describe the time evolution of the
variables and the system they represent$^{9-11}$. 
\par 
A solution of the dynamical system is a vector function 
${\bf z}({\bf z}_{0},t)$, that satisfies (10) and the initial
condition
\beq
{\bf z}({\bf z}_{0},0)={\bf z}_{0} \; .  
\eeq
Usually one writes simply ${\bf z}(t)$ without the initial condition
dependence.
\par 
The time evolution of ${\bf z} \in \Omega$ is obtained with the one
parameter group of diffeomorfism $g^{t}$: $\Omega \to \Omega$, such that 
\beq
{d\over dt}(g^{t}{\bf z})|_{t=0}={\bf f} ({\bf z},0) \; . 
\eeq
The group $g^{t}$ is called phase flux and the solution is 
called orbit. The system is called Hamiltonian, 
if the dimension of $\Omega$ is even and there exists a function 
$H({\bf z},t)$ given by
\beq
{\bf f} ({\bf z}(t),t)=J\nabla H({\bf z},t) \; ,
\eeq
where:
\beq
J= \left( \matrix {0 & I \cr -I & 0 \cr } \right)  
\eeq
is the symplectic matrix and $H({\bf z},t)$ is the Hamiltonian function. 
\par 
On the phase space $\Omega$ one usually defines a probability measure 
$\mu : \Omega \to \Omega$, such that $\mu (\Omega )=1$. 
If we choose a subspace A of 
$\Omega$, the system is measure preserving if 
\beq
\mu (g^{t}A)=\mu (A) \; . 
\eeq
We observe that for measure preserving dynamical systems 
one gets $\hbox{div}({\bf f})=0$.  
It is well known that Hamiltonian systems preserve their measure: 
the Liouville measure. Dynamical systems which do not preserve their 
measure are called dissipative, and usually have 
a measure contraction in time evolution. 
\par 
The dynamics of a system is called regular if the orbits are
stable to infinitesimal variations of initial conditions. It 
is called chaotic if the orbits are
unstable to infinitesimal variations of initial
conditions. Useful quantities to calculate this behaviour are the
Lyapunov exponents, which give the stability of a single orbit. 
\par
A vector of the tangent space $T\Omega_{\bf z}$ to the phase space 
$\Omega$ in the position ${\bf z}$ is given by
\beq
{\bf \omega} ({\bf z})=\lim_{s \to 0} {{\bf r}(s)-{\bf r}(0) \over s} \; ,
\eeq
where ${\bf r}(0)={\bf z}$ and ${\bf r}(s) \in \Omega$. The tangent
space vectors are the velocity vectors of the curves on M; there are
obviously N independent vectors.
\par
Now we can define the Lyapunov exponent
\beq
\lambda ({\bf z})=\lim_{t\to \infty}{1\over t}\ln{|\omega (t)|}, 
\eeq
where ${\bf \omega} (t)$ is a tangent vector to ${\bf z}(t)$ with the
condition that $|{\bf \omega} (0)|=1$.
\par 
It can be demonstrated that the limit given by the previous equation 
exists for a
compact phase space, and that it is metric independent. Fixing an
orbit in the N dimensional phase space, there are N distinct
exponents $\lambda_{1},...,\lambda_{N}$, called first order
Lyapunov exponents. If the orbit has positive Lyapunov exponents, it
is chaotic.
\par 
To characterize globally the chaoticity of a system, we can 
introduce the Kolmogorov--Sinai entropy, which is given by 
\beq
h_{KS}(\mu )=\int_{A} d\mu ({\bf z}) \sum_{\lambda_{i}>0} 
\lambda_{i}({\bf z}) \; ,
\eeq
with $A$ subspace of $\Omega$ and $\lambda_{i}$ Lyapunov exponents. 
The Kolmogorov--Sinai entropy is a very useful tool 
for showing chaotic behaviour in the region $A$. 
\par 
A system is called ergodic if the time average is equal to
phase space average
\beq
\lim_{t\to\infty}{1\over t}\int_{0}^{t} dt f(g^{t}{\bf z}(t))=
\int_{\Omega} d\mu ({\bf z}) f({\bf z}) \; . 
\eeq
Incidentally, as is well known, Boltzmann started from the ``ergodic
hypothesis" to obtain statistical mechanics of equilibrium. 
But ergodicity is not sufficient to reach an equilibrium state:
one must consider mixing systems.
\par 
In a mixing system, every finite element of the phase space
occupies for $t\to \infty$ the entire phase space $\Omega$; 
more precisely: 
$\forall A,B \subset \Omega$ with $\mu(A)$ and $\mu (B) \not= 0$,
\beq
\lim_{t \to \infty} {\mu (B \cap g^{t}A)\over \mu (B)}=\mu (A) \; .
\eeq
To have quantitative information of orbit separations, we must
introduce K--systems (Kolmogorov), which are mixing systems
with a positive metric entropy, i.e. $h_{KS}>0$. 
Such systems are typical chaotic systems. 
  
Among the K--systems, the most unpredictable ones are the
B--systems (Bernoulli), which have the Kolmogorov--Sinai
entropy equal to the entropy of every partition, i.e. 
$h_{KS}=h(A_{i}(0),\mu )$, $\forall A_{i}(0)$. 

\subsection{Hamiltonian dynamics}
\par
Let us consider an Hamiltonian system with $n$ degrees of freedom 
described by the Hamiltonian function $H({\bf z})$, where 
${\bf z}=(q_1,...q_n,p_1,...,p_n)$ so that the phase space is 
$N=2n$ dimensional. The Hamiltonian system 
is called integrable if there are $N$ functions 
$F_{i}=F_{i}({\bf z})$ defined on $\Omega$ in involution:
\beq
[ F_{i},F_{j} ]_{PB}=\sum_{k=1}^{n} 
{\partial F_i\over \partial q_k}{\partial F_j\over \partial p_k}-
{\partial F_j\over \partial q_k}{\partial F_i\over \partial p_k}
=0, \quad \forall i,j 
\eeq
and linearly independent. $[\; ,\; ]_{PB}$ are the Poisson brackets.
\par
For conservative systems we have $F_{1}=H({\bf z})$ and also 
\beq
{dF_{i}\over dt}=[H,F_{i}]_{PB}=0 \; . 
\eeq
Because there are n constants of motion, 
every orbit can explore only the $n$ dimensional manifold 
$\Omega_{f}=\{{\bf z}: F_{i}({\bf z})=f_{i}, i=1,...,n\}$. 
If $\Omega_{f}$ is compact and connected, it is equivalent to a n
dimensional torus $T^{n}=\{ (Q_{1},...,Q_{n})\quad mod\quad 2\pi \}$. 
There are n irreducible and independent circuits $\gamma_{i}$ on 
$\Omega_{f}$ and there exists a canonical transformation 
$({\bf p},{\bf q}) \to ({\bf P},{\bf Q})$, 
generated by the function $S({\bf q},{\bf P})$, such that 
\beq
\ba{ccc}
P_{i} & = & \oint_{\gamma_{i}} d{\bf q}\cdot{\bf p} 
\\ 
Q_{i} & = & {\partial S\over \partial P_{i}} \; . 
\ea
\eeq
The $P_{i}$ are called action variables and the $Q_{i}$ are
called angle variables. 
The moments ${\bf p}$ and coordinates ${\bf q}$ are periodic 
functions of ${\bf Q}$ with period $2\pi$.  
The Hamiltonian depends only on action variables, i.e. 
$H=H({\bf P})$. 
\par
Adding a small perturbation $V({\bf P},{\bf Q})$ 
to an integrable Hamiltonian $H_0({\bf P})$, the total Hamiltonian can 
be written:
\beq
H({\bf P},{\bf Q})=H_{0}({\bf P})+g V({\bf P},{\bf Q}) \; ,
\eeq
and, generically, the integrability is destroyed. As a consequence, 
parts of phase space become
filled with chaotic orbits, while in other parts the toroidal
surfaces of the integrable system are deformed but not destroyed; thus 
we have a quasi--integrable system. 
By growing $g$, chaotic motion develops near the 
regions of phase space where all the frequencies on the torus 
$\omega_{i}={\partial H({\bf P}) \over \partial P_{i} }$ are
commensurate. Conversely, tori of the integrable system,  
on which the $\omega_{i}$ are incommensurate, 
are deformed, but not destroyed immediately 
(Kolmogorov--Arnold--Moser (KAM) theorem)$^{9,18}$. 
As $g$ increases, the phase space generically
develops a highly complex structure, with islands of regular
motion (filled with quasi--periodic orbits) 
interspersed in regions of chaotic motion, but containing in turn
more regions of chaos. As $g$ grows further, the
fraction of phase space filled with chaotic orbits grows until it
reaches unity as the last KAM surface is destroyed. Then the
motion is completely chaotic everywhere, except possibly for
isolated periodic orbits$^{9,18}$. 
\par 
It is very useful to plot a $2n-1$ surface of section ${\cal P} 
\subset \Omega$, called Poincar\'e section. 
For an integrable system with two degrees of 
freedom, the $q_1=0$ Poincar\'e section of a rational (resonant) 
torus is a finite number of points along a closed curve, 
while the section of an irrational
(non resonant) torus is a continuous closed curve. 
Adding a perturbation, the section presents closed curves (KAM tori), 
whose points are stable (elliptic), and also curves formed
by substructures, residua of resonant tori, whose points are
unstable (hyperbolic). As the perturbation parameter increases, the
closed curves are distorted and reduced in number.

\subsection{Quantum Chaos}
\par
We use the term quantum chaotic system in the precise and 
restricted sense of a quantum system whose classical analogue 
is chaotic. In particular we concentrate on energy levels 
of quantum systems (see, for example, Ref. 18 and 19). 
\par 
Let us consider a classical regular Hamiltonian system. 
The short--range properties of the corresponding quantal 
spectrum tend to resemble those of a spectrum of randomly distributed 
numbers. This is because regular classical motion is 
associated with integrability or separability of the classical equations of 
motion. In quantum mechanics the separability corresponds to a number of 
independent conserved quantities (such as angular momentum), and each 
energy level can be characterised by the associated quantum numbers. 
Superimposing the terms arising from the 
various quantum numbers, a spectrum is generated like that of 
random numbers, at least over short intervals. In particular, 
the distribution $P(s)$ of nearest--neighbour spacings 
$s_i=(\epsilon_{i+1}-\epsilon_i)/d$, where $d$ is the mean level spacing,  
is expected to follow the Poisson limit, i.e. $P(s)=\exp{(-s)}$.  
\par
Instead, when the classical dynamics of a physical system is chaotic, 
the system cannot be integrable 
and there must be fewer constants of motion than degrees of freedom. 
Quantum mechanically this means that once all good quantum numbers due to 
obvious symmetries etc. are accounted for, the energy levels cannot simply 
be labelled by quantum numbers associated with certain constants of motion. 
The short--range properties of the energy spectrum then tend to resemble 
those of eigenvalue spectra of matrices with randomly chosen elements and 
one gets a result very close to 
$P(s)= (\pi / 2) s \exp{(-{\pi \over 4}s^2)}$, which is the 
so--called Wigner distribution. 
\par
The distribution $P(s)$ is the best spectral statistics to analyze 
shorter series of energy levels and 
the intermediate regions between order and chaos. 
This distribution can be compared to the Brody distribution 
\beq
P(s,\omega)=\alpha (\omega +1) s^{\omega} \exp{(-\alpha s^{\omega+1})} \; ,
\eeq
with 
\beq
\alpha = \big( \Gamma [{\omega +2\over \omega+1}] \big)^{\omega +1} \; .
\eeq
The Brody distribution 
interpolates between the Poisson distribution ($\omega =0$) 
of integrable systems and the Wigner distribution ($\omega =1$) of 
chaotic ones, and thus the parameter $\omega$ can be used as a simple 
quantitative measure of the degree of chaoticity. 

\section{A model for inflationary cosmology}
\par
In this Section we study the stability of a scalar inflaton 
field$^{12,13}$ and analyze its bifurcation properties 
in the framework of the dynamical system theory. 
\par
It is generally believed that the universe, at a very early stage after 
the big bang, exhibited a short period of exponential expansion, the 
so--called inflationary phase. 
In fact the assumption of an inflationary universe 
solves three major cosmological problems: the flatness problem, the 
homogeneity problem, and the formation of structure problem$^{3}$. 
\par
The Friedmann--Robertson--Walker metric of a homogeneous and isotropic 
expanding universe is given by
\beq
ds^2=dt^2-a^2(t)\big[ {dr^2\over 1 - k r^2}+r^2(d\theta^2 +
\sin^2{\theta} d\varphi^2 )\big],
\eeq
where $k=1,-1$, or $0$ for a closed, open, or flat universe, and $a(t)$ 
is the scale factor of the universe. 
\par
The evolution of the scale factor $a(t)$ is given by 
the Einstein equations
$$
{\ddot a}=-{4\pi \over 3}G(\rho + 3 p)a , 
$$
\beq
\big({{\dot a} \over a}\big)^2+{k\over a^2}=
{8\pi \over 3}G \rho ,
\eeq
where $\rho$ is the energy density of matter in the universe, 
and $p$ its pressure. The gravitational constant $G=M_p^{-2}$ 
(with $\hbar =c=1$), 
where $M_p =1.2 \cdot 10^{19}$ GeV is the Plank mass, and 
$H_u={\dot a}/a$ is the Hubble "constant", which in general is a function 
of time. 
\par
The inflationary models postulate the existence of a scalar field 
$\phi$, the so--called inflation field, with Lagrangian  
\beq
L={1\over 2}\partial_{\mu}\phi \partial^{\mu}\phi -V(\phi )
\eeq
where the potential $V(\phi )$ depends on the type of inflation model 
considered. The scalar field, 
if minimally coupled to gravity, satisfies the equation
\beq
\Box \phi = {\ddot \phi} + 3 \big({\dot a\over a}\big) 
{\dot \phi} - {1\over a^2} \nabla^2 \phi 
= -{\partial V \over \partial \phi},
\eeq
where $\Box$ is the covariant d'Alembertian operator. 
The Hubble "constant" $H_u$ is related to the energy density of 
the field by 
\beq
H_u^2 +{k\over a^2}=
\big({{\dot a} \over a}\big)^2+{k\over a^2}=
{8 \pi G \over 3} \big[
{{\dot \phi}^2 \over 2}+{(\nabla \phi )^2\over 2}+ V(\phi ) 
\big].
\eeq
\par
In a flat universe $k=0$ and, if the inflaton field is 
sufficiently uniform (i.e. ${\dot \phi}^2$, $(\nabla \phi )^2 << V(\phi )$), 
we obtain an homogenous field theory in one dimension
\beq
{\ddot \phi } + 3 H_u(\phi ) {\dot \phi} + 
{\partial V \over \partial \phi}=0,  
\eeq
where the Hubble "constant" $H_u$ is an explicit function of $\phi$: 
\beq
H_u^2 = {8 \pi G \over 3} V(\phi ).
\eeq

\subsection{Local instability for the inflationary self--energy}
\par
The second order equation of motion of our cosmological model can be 
written as a system of two first order differential equations 
\beq
{\dot {\bf z}}={\bf f}({\bf z})
\eeq
where ${\bf z}=(\phi , \chi )$ is a point in the two dimensional 
phase space and ${\bf f}=(f_1,f_2)$ is given by
\beq
f_1(\phi ,\chi )=\chi ,
\quad 
f_2(\phi ,\chi )= - 3 H_u(\phi ) \chi 
- {\partial V(\phi )\over \partial \phi} .
\eeq
The system is non--conservative because the function
\beq
\hbox{div}({\bf f})=
{\partial g_1 \over \partial \phi }+{\partial g_2 \over \partial \chi} 
=- 3 H_u(\phi ) 
\eeq
is not identically zero. The fixed points of the system are those for which 
$f_1(\phi ,\chi )=0$ and $f_2(\phi ,\chi )=0$, i.e 
\beq 
\chi =0, 
\quad 
{\partial V(\phi )\over \partial \phi}=0 .
\eeq 
\par
The deviation $\delta {\bf z}(t)={\hat {\bf z}}(t)-{\bf z} (t)$ 
from the two initially neighboring trajectories 
${\bf x}$ and ${\hat{\bf x}}$ in the phase space 
satisfies the linearized equations of motion
\beq
{d \over dt}\delta {\bf z}(t)= \Gamma (t) \delta {\bf z}
\eeq
where $\Gamma (t)$ is the stability matrix
\beq
\Gamma (t)= {\left(\matrix {0 & 1 \cr 
- {\partial^2 V\over \partial \phi^2} - 3\chi {\partial H\over \partial \phi}
& - 3H_u(\phi ) \cr } \right) }.
\eeq
At least if an eigenvalue of $\Gamma (t)$ is real 
the separation of the trajectories grows exponentially 
and the motion is unstable. 
Imaginary eigenvalues correspond to stable motion. 
In the limit of time that goes to infinity, from 
the eigenvalues of the stability 
matrix we can obtain the Lyapunov exponents. For 
two--dimensional dynamical system the Lyapunov exponents can not be 
positive$^{9}$ and so the system is not chaotic, 
i.e. there is not global instability. 
However, we can be assured that the universe is crowded with many 
interacting fields of which the inflaton is but one. 
The nonlinear nature of these interactions can result in a complex 
chaotic evolution of the universe and the local instability 
of the inflaton field is a precursor phenomenon of chaotic motion. 
\par
The eigenvalues of the stability matrix are given by
\beq
\sigma_{1,2}=-{3\over 2}H_u(\phi )\pm {1\over 2}\sqrt{9 H_u^2(\phi ) 
-4 {\partial^2 V\over \partial \phi^2} 
-12 \chi {\partial H_u \over \partial \phi} }.
\eeq
The pair of eigenvalues become real and there 
is exponential separation of neighboring trajectories, i.e.  
unstable motion, if
\beq
{\partial^2 V \over \partial \phi^2} + 
3\chi {\partial H_u\over \partial \phi}< 0.
\eeq
Particularly when $\chi =0$, e.g. the fixed points, we obtain 
local instability when
\beq
{\partial^2 V \over \partial \phi^2} < 0,
\eeq
i.e. for negative curvature of the potential energy. The fixed points 
are stable if they are point of local minimum of $V(\phi )$ and unstable 
if are points of local maximum. 
\par
The potential $V(\phi )$ depends on the 
type of inflation model considered, and it is usually some 
kind of double--well potential. We choose a symmetric double--well 
potential
\beq
V(\phi )={\lambda \over 4}(\phi^2 -v^2)^2 ,
\eeq
where $\pm v$ are the values of the inflaton field in the vacuum, 
i.e. the points of minimal energy of the system. 
\par
We observe that the inflaton field value in the vacuum $v$ is a 
bifurcation parameter. Bifurcation is used to indicate a qualitative 
change in the features of the system under the variation of one or more 
parameters on which the system depends. 
First of all we consider the case $v=0$, i.e. $V(\phi )=(\lambda /4) \phi^4$. 
In this situation there is only one fixed point $(\phi^*=0,\chi^*=0)$ 
which is a stable one being 
\beq
{\partial^2 V\over \partial \phi^2}=3\lambda \phi^2 \geq 0. 
\eeq
The fixed point $(\phi^*=0,\chi^*=0)$ is a point attractor.
\par
Instead for $v\neq 0$ there are three fixed points 
\beq
(\phi^* =0,\chi^* =0), \quad (\phi^* = v,\chi^* =0) ,
\quad (\phi^* =-v,\chi^* =0),
\eeq
and the condition for the instability becomes
\beq
-{v\over \sqrt{3}}< \phi < {v\over \sqrt{3}}.
\eeq
Obviously $(\phi^* =0,\chi^* =0)$ is an unstable fixed point, and 
in particular a saddle point because the stability matrix has 
real and opposite eigenvalues. 
On the other hand $(\phi^* =\pm v, \chi^* =0)$ are stable 
fixed points. 
\par
There are four possible functions for the Hubble "constant" 
\beq
H_u(\phi )= \pm \gamma |\phi^2-v^2| , 
\eeq
but also 
\beq
H_u(\phi )= \pm \gamma (\phi^2-v^2) , 
\eeq
where $\gamma=\sqrt{2\pi G\lambda / 3}$ is the dissipation parameter. 
The choice of the Hubble function is crucial for the dynamical 
evolution of the system. 
\par
In certain non--conservative systems 
we could find closed trajectories or limit 
cycles toward which the neighboring trajectories spiral on both sides. 
It is sometime possible to know that no limit cycle exist and 
the Bendixson criterion$^{22}$, which establishes a condition for the 
non--existence of closed trajectories, is useful in some cases. 
Bendixson criterion is as follows: if $\hbox{div}({\bf f})$ is not zero and 
does not change its sign within a domain $D$ of the phase space, 
no closed trajectories can exist in that domain. 
In our case we have $\hbox{div}({\bf f})=- 3 H_u(\phi )$, 
and so the presence of periodic orbit is related to the sign of $H_u(\phi )$. 
\par
If $H_u(\phi )=\gamma |\phi^2-v^2|$ we do not find periodic orbits 
and the inflaton field goes to one of its two stable fixed points, 
which are points attractors (see Figure 1). The vacuum is degenerate but 
if we choose an initial condition around the saddle point 
there is a dynamical symmetry breaking towards the positive $v$ 
or negative $-v$ value of the inflaton field in the vacuum. 
This symmetry breaking is unstable because neighbour 
initial conditions can go in different point attractors. 
\par
Instead, if we choose $H_u(\phi )=\gamma (\phi^2-v^2)$ the numerical 
calculations of Figure 2 show that exists a limit cycle, 
the two stable fixed points are not point attractors, 
and the inflaton field oscillates forever. 
Obviously more large is $v$ more large is the limit cycle. 
 
\subsection{A limit cycle in the cosmological model}
\par
The equation of motion of the inflaton field with 
$H_u(\phi )=\gamma (\phi^2-v^2)$ reads 
\beq
{\ddot \phi}+ 3 \gamma (\phi^2 - v^2) {\dot \phi}+ 
\lambda \phi ( \phi^2 - v^2 ) = 0 \; .
\eeq
This equation can be written as
\beq
{d\over dt}\big[ {\dot \phi} + 3 \gamma \int_0^{\phi}(u^2-v^2) du \big] 
+ \lambda \phi (\phi^2 - v^2) = 0 \; ,
\eeq
and if we put
\beq
F(\phi )= 3 \int_0^{\phi}(u^2-v^2) du = \phi (\phi^2 - 3 v^2) \; , 
\;\;\;\; G(\phi ) = \phi (\phi^2 - v^2 ) \; ,
\eeq
and also $\omega = {\dot \phi} + \gamma F(\phi )$, we obtain the system
\beq
\ba{ccc}
{\dot \phi} & = & \omega - \gamma F(\phi ) 
\\
{\dot \omega} & = & - \lambda G(\phi ) \; . 
\ea  
\eeq 
For systems of this kind the Lienard theorem$^{23}$ states that there 
is an unique and stable limit cycle 
if the following conditions are satisfied: 
$F(\phi )$ is an odd function and $F(\phi )=0$ only at $\phi =0$ and 
$\phi = \pm \alpha $; $F(\phi ) < 0$ for $0< \phi < \alpha $, 
$F(\phi ) > 0$ and is increasing for $\phi > \alpha $; $G(\phi )$ is an 
odd function and $\phi G(\phi ) >0 $ for all 
$\phi > \alpha$. It is easy to check that the functions $F(\phi )$ and 
$G(\phi )$ satisfy all the conditions of the Lienard 
theorem with $\alpha = v$. The cubic force $G(\phi )$ tends to reduce any 
displacement for large $|\phi |$, whereas the damping $F(\phi )$ is 
negative at small $|\phi |$ and positive at large $|\phi |$. 
Since small oscillations are pumped up and 
large oscillations are damped down, 
it is not surprising that the system tends to seattle into a self--sustained 
oscillation of some intermediate amplitude.  
\par
Let us consider a typical trajectory of the system. 
After the scaling $\psi = \lambda \omega $ we obtain 
\beq
\ba{ccc}
{\dot \phi} & = & \lambda [\psi - {\gamma\over \lambda }F(\phi )]  \; , 
\\
{\dot \psi} & = & - G(\phi ) \; .
\ea 
\eeq
The cubic nullcline $\psi = (\gamma /\lambda ) F(\phi ) $ is the key to 
understand the motion. Suppose that $\lambda >>1$ and 
the initial condition is far from 
the cubic nullcline, then we have 
$|{\dot \phi }|\sim O(\lambda ) >> 1$; hence the velocity 
is enormous in the horizontal direction and tiny in the vertical direction, 
so trajectories move practically horizontally. If the initial condition 
is above the nullcline then ${\dot \phi} >0$, thus the trajectory moves 
sideways toward the nullcline. However, once the trajectory 
gets so close that 
$\psi \simeq (\lambda /\gamma) F(\phi )$ then 
the trajectory crosses the nullcline vertically and moves slowing along 
the backside of the branch until it reaches the knee and can jump sideways 
again. The period $T$ of the limit cycle is essentially the time required 
to travel along the two slow branches, since the time spent in the jumps is 
negligible for large $\lambda$. 
By symmetry, the time spent on each branch is the same so we have 
\beq
T \simeq 2 \int_{t_A}^{t_B} dt \; 
\eeq
where $A$ and $B$ are the initial and final 
points on the positive slow branch. 
To derive an expression for $dt$ we note that on the slow branches with 
a good approximation $\psi \simeq (\gamma /\lambda ) F(\phi )$ and thus 
\beq
{d\psi \over dt} \simeq {\gamma \over \lambda } F'(\phi ) {d\phi \over dt} 
= 3 {\gamma \over \lambda} (\phi^2 - v^2){d\phi \over dt} \; .
\eeq
Since $d\psi / dt = - \phi (\phi^2 -v^2)$, we obtain 
$dt \simeq - 3 {\gamma\over \lambda } {d\phi \over \phi }$, 
on the slow branches. The slow positive branch begins at 
$\phi_A = 2\gamma v/\lambda$ and ends at $\phi_B=\gamma v/\lambda$. 
Because $\gamma = \sqrt{2\pi G \lambda /3}$ we get 
$T\simeq 2 \ln{2} \sqrt{6\pi G\over \lambda}$. 

\section{The homogenous SU(2) YMH system}
\par
Now we study the suppression of classical chaos 
in the spatially homogenous SU(2) Yang--Mills--Higgs (YMH) system 
induced by the Higgs field$^{4,14,15}$. 
We analyze also the energy fluctuation properties 
of the system, which give a clear quantum signature 
of the classical chaos--order transition of the system. 
\par 
The SU(2) YMH system describes the interaction between a scalar Higgs 
field $\phi$ and three non--Abelian Yang--Mills fields $A_{\mu}^a$, 
$a=1,2,3$. The Lagrangian density of the YMH system is given by 
\beq
L={1\over 2}(D_{\mu}\phi )^+(D^{\mu}\phi ) -V(\phi ) 
-{1\over 4}F_{\mu \nu}^{a}F^{\mu \nu a} \; ,
\eeq
where
\beq
(D_{\mu}\phi )=\partial_{\mu}\phi - i g A_{\mu}^b T^b\phi 
\; ,
\eeq
\beq
F_{\mu \nu}^{a}=\partial_{\mu}A_{\nu}^{a}-\partial_{\nu}A_{\mu}^{a}+
g\epsilon^{abc}A_{\mu}^{b}A_{\nu}^{c} \; ,
\eeq
with $T^b=\sigma^b/2$, $b=1,2,3$, generators of the SU(2) algebra, 
and where the potential of the scalar field (the Higgs field) is
\beq
V(\phi )=\mu^2 |\phi|^2 + \lambda |\phi|^4 \; .
\eeq
We work in the (2+1)--dimensional Minkowski space ($\mu =0,1,2$) and 
choose spatially homogeneous Yang--Mills and the Higgs fields
\beq
\partial_i A^a_{\mu} = \partial_i \phi = 0 \; , \;\;\;\; i=1,2
\eeq
i.e. we consider the system in the region in which space fluctuations of 
fields are negligible compared to their time fluctuations. 
\par
In the gauge $A^a_0=0$ and using the real triplet representation for the 
Higgs field we obtain
$$
L={\dot{  \phi}}^2 +
{1\over 2}({\dot {  A}}_1^2+{\dot {  A}}_2^2) 
-g^2 [{1\over 2}{  A}_1^2 {  A}_2^2 
-{1\over 2} ({  A}_1 \cdot {  A}_2)^2+
$$
\beq
+({  A}_1^2+{  A}_2^2){  \phi}^2 
-({  A}_1\cdot {  \phi})^2 -({  A}_2 \cdot {  \phi})^2 ]
-V( {  \phi} ) \; ,
\eeq
where ${  \phi}=(\phi^1,\phi^2,\phi^3)$, 
${  A}_1=(A_1^1,A_1^2,A_1^3)$ and ${  A}_2=(A_2^1,A_2^2,A_2^3)$. 
\par
When $\mu^2 >0$ the potential $V$ has a minimum at $|{  \phi}|=0$, 
but for $\mu^2 <0$ the minimum is at 
$$
|{  \phi}_0|=\sqrt{-\mu^2\over 4\lambda }=v \; ,
$$
which is the non zero Higgs vacuum. This vacuum is degenerate 
and after spontaneous symmetry breaking the physical vacuum can be 
chosen ${  \phi}_0 =(0,0,v)$. If $A_1^1=q_1$, $A_2^2=q_2$ 
and the other components of the Yang--Mills fields are zero, 
in the Higgs vacuum the Hamiltonian of the system reads 
\beq
H={1\over 2}(p_1^2+p_2^2)
+g^2v^2(q_1^2+q_2^2)+{1\over 2}g^2 q_1^2 q_2^2 \; ,
\eeq
where $p_1={\dot q_1}$ and $p_2={\dot q_2}$. Here $w^2=2 g^2v^2$ is the 
mass term of the Yang--Mills fields. This YMH Hamiltonian is 
a toy model for classical non--linear dynamics, with the attractive feature 
that the model emerges from particle physics. 

\subsection{From chaos to order in the YMH system}

The chaotic behaviour of the YMH system can be studied by using 
the Toda criterion of the Gaussian curvature 
of the potential energy$^{20,21}$. 
For our YMH system the potential energy is given by 
\beq
V(q_1 ,q_2)=g^2v^2(q_1^2+q_2^2)+{1\over 2}g^2 q_1^2 q_2^2 \; .
\eeq 
At low energy, the motion near the minimum of the potential, 
where the Gaussian curvature is positive, 
is periodic or quasi--periodic and is 
separated from the instability region by a line of zero curvature; 
if the energy is increased, the system will be, for some initial conditions, 
in a region of negative curvature, where the motion is chaotic. 
According to this scenario, the energy $E_c$ of chaos--order transition 
is equal to the minimum value of the line of zero Gaussian 
curvature $K_G(q_1 ,q_2 )$ on the potential--energy surface. 
For our potential the gaussian curvature vanishes at the points 
that satisfy the equation
\beq
(2g^2v^2 +g^2 q_2^2)(2g^2v^2+g^2q_1^2)-4g^4 q_1^2q_2^2=0 \; .
\eeq
It is easy to show that the minimal energy on the 
zero--curvature line is given by: 
\beq
E_c=V_{min}(K_G=0,\bar{q_1})=6 g^2 v^4 \; , 
\eeq
and by inverting this equation 
we obtain $v_c=(E /6g^2)^{1/4}$. Thus the curvature 
criterion suggest that there is a order--chaos transition 
by increasing the energy $E$ 
of the system and a chaos--order transition by increasing 
the value $v$ of the Higgs field in the vacuum. 
Thus, there is only one transition regulated 
by the unique parameter $E/(g^2v^4)$. 
\par 
It is important to stress that 
the Toda criterion is not a fully reliable indicator of chaos$^{21}$. 
In fact, the local instability of the Toda Criterion 
does not necessarily imply the global one and the idea 
of an order--chaos transition with a critical energy is not 
strictly correct. 
The Toda curvature criterion should therefore be combined 
with the Poincar\`e sections, 
which are shown in Figure 3. The numerical results 
confirm the analytical predictions of the curvature criterion: 
with $E =10$ and $g=1$ we get the critical value of the onset of chaos 
$v_c=(E /6g^2)^{1/4}\simeq 1.14$. 

\subsection{Spectral statistics of the YMH system}
\par
In quantum mechanics the generalized coordinates of the YMH system 
satisfy the usual commutation rules $[{\hat q}_k,{\hat p}_l]=i\delta_{kl}$, 
with $k,l=1,2$. Introducing the creation and destruction operators
\beq
{\hat a}_k=\sqrt{\omega \over 2}{\hat q}_k + 
i \sqrt{1\over 2\omega}{\hat p}_k \; ,
\;\;\;\;
{\hat a}_k^+ = \sqrt{\omega \over 2}{\hat q}_k - 
i \sqrt{1\over 2\omega}{\hat p}_k \; ,
\eeq
the quantum YMH Hamiltonian can be written 
\beq
{\hat H}={\hat H}_0 + {1\over 2} g^2 {\hat V} \; ,
\eeq
where
\beq
{\hat H}_0= \omega ({\hat a}_1^+ {\hat a}_1 + {\hat a}_2^+ {\hat a}_2 + 1) \; ,
\eeq
\beq
{\hat V}= {1 \over 4 \omega^2} ({\hat a}_1 +{\hat a}_1^+)^2 
({\hat a}_2 +{\hat a}_2^+)^2 \; ,
\eeq
with $\omega^2 = 2 g^2 v^2$ and $[{\hat a}_k,{\hat a}_l^+] = \delta_{kl}$, 
$k,l=1,2$. 
\par
We compute the energy levels of the YMH system with 
a numerical diagonalization of the truncated matrix of the quantum 
YMH Hamiltonian in the basis of the harmonic oscillators 
(see also Ref. 16 and 17). If $|n_1 n_2>$ is the basis of the 
occupation numbers of the two harmonic oscillators, the matrix elements are
\beq
<n_{1}^{'}n_{2}^{'}|{\hat H}_0|n_{1}n_{2}>= \omega (n_1+n_2+1) 
\delta_{n_{1}^{'}n_{1}} \delta_{n_{2}^{'}n_{2}} \; ,
\eeq
and
$$
<n_{1}^{'}n_{2}^{'}|{\hat V}|n_{1}n_{2}>=
{1 \over 4 \omega^2}
[\sqrt{n_{1}(n_{1}-1)} \delta_{n^{'}_{1}n_{1}-2}
+\sqrt{(n_{1}+1)(n_{1}+2)}\delta_{n^{'}_{1}n_{1}+2}+
$$
\beq
+(2n_{1}+1)\delta_{n^{'}_{1}n_{1}}]
\times[\sqrt{n_2 (n_2-1)}\delta_{n^{'}_2 n_2-2}+ \sqrt{(n_2+1)(n_2+2)}
\delta_{n^{'}_2 n_2+2}+ (2n_2+1)\delta_{n^{'}_2 n_2}] \; .
\eeq
The symmetry of the potential enables us to split 
the Hamiltonian matrix into 4 sub--matrices 
reducing the computer storage required. These sub--matrices are related 
to the parity of the two occupation numbers $n_1$ and $n_2$: 
even--even, odd--odd, even--odd, odd--even. 
The numerical energy levels depend on the dimension of the truncated matrix: 
we compute the numerical levels in double precision 
increasing the matrix dimension until the first 100 levels converge 
within $8$ digits (matrix dimension $1156\times 1156$). 
\par
We have seen previously that the most used quantity 
to study the local fluctuations of the energy levels 
is the distribution $P(s)$ of nearest--neighbour spacings 
$s_i$ of the energy levels. 
It is obtained by accumulating the number of spacings that lie within 
the bin $(s,s+\Delta s)$ and then normalizing $P(s)$ to unit. 
\par
We use the first $100$ energy levels of the 4 sub--matrices 
to calculate the $P(s)$ distribution. 
In order to remove the secular variation of the level density as a function 
of the energy $E$, for each value of the coupling constant the 
corresponding spectrum is mapped into one which has a constant level density. 
\par
The Figure 4 shows the $P(s)$ distribution of Brody 
for three different values of the Higgs vacuum $v$. The best fit Brody 
parameter $\omega$ is obtained by using the nearest--neighbour
spacings of the first $100$ unfolded energy levels of the YMH system. 
There is Wigner--Poisson transition by increasing the value $v$ 
of the Higgs field in the vacuum. Thus, by using the P(s) distribution, 
it is possible to give a quantitative measure 
of the degree of quantal chaoticity of the system. 
Our numerical calculations show clearly the quantum 
chaos--order transition and its correspondence to the classical one.  

\section{Conclusions}
\par
We have seen that spatially homogeneous field theories 
can be studied as dynamical systems. After a brief review of 
the dynamical system theory, we have discussed two 
schematic models of field theory. 
\par
First, we have considered the stability of a non--conservative scalar 
inflaton field. 
The value of the inflaton field in the vacuum is a bifurcation parameter 
which changes dramatically the phase space structure. 
The main point is that for some functional solutions of the Hubble "constant" 
the system goes to a limit cycle, i.e. to a periodic orbit. 
The inflaton field is not chaotic but its 
local instability can give rise to a complex chaotic evolution of the 
universe due to its nonlinear interactions with other fields. 
In the future it will be very interesting to study these effects 
which can perhaps lead to some observable implications 
like a fractal pattern in the spectrum of density fluctuations. 
\par
We have then analyzed the non--Abelian SU(2) Yang--Mills--Higgs 
system.  We have given an analytical estimation (confirmed by 
numerical results of Poincar\`e sections) 
of the classical chaos--order transition 
as a function of the Higgs vacuum, the Yang--Mills coupling constant 
and the energy of the system. 
A quantum signature of a chaos--order transition 
has been obtained 
by using the distribution $P(s)$ of nearest--neighbour spacings. 
The Wigner--Poisson transition of the $P(s)$ distribution follows very well 
the classical results of the Poincar\`e sections. 
\par
To conclude, we observe that there are yet many open problems about chaos in 
field theory. We make a list of some of them: 
i) spatial chaos and  
space--time chaos; ii) classical and quantum chaos in 
more realistic systems, for example in QCD (some 
results can be found in Ref. 7 and 8); iii) connection between chaos 
and critical phenomena (finite temperature field theory). 

\newpage

\parindent=0.pt
\section*{Figure Captions}
\vspace{0.6 cm}

{\bf Figure 1}: The Hubble function {\it vs} time (top) and the phase space 
trajectory of the inflaton field (bottom); for $H_u(\phi )= 
\gamma |\phi^2 - v^2 |$ with $\gamma =1/2$, $\lambda =3$ and $v=1$. 
Initial conditions: $\phi =0$ and ${\dot \phi}=1/2$. 

{\bf Figure 2}: The Hubble function {\it vs} time (top) and the phase space 
trajectory of the inflaton field (bottom); for $H_u(\phi )= 
\gamma (\phi^2 - v^2 )$ with $\gamma =1/2$, $\lambda =3$ and $v=1$. 
Initial conditions: $\phi = -1/2$ and ${\dot \phi}=0$. 

{\bf Figure 3}: The Poincar\`e sections of the YMH system. From the top: 
$v=1$, $v=1.1$ and $v=1.2$. Energy $E = 10$ and interaction $g=1$. 

{\bf Figure 4}: $P(s)$ distribution of Brody of the YMH system. 
First $100$ energy levels and $g=1$. The best fit Brody parameter 
is given by: $\omega=0.92$ for $v=1.0$, $\omega =0.34$ for $v=1.1$ 
and $\omega =0.01$ for $v=1.2$.

\newpage

\section*{References}

$^{1}$A.L. Fetter and J.D. Walecka, {\it Quantum Theory 
of Many--Particle Systems} (McGraw--Hill, New York, 1971) 
\\
$^{2}$C. Itzykson and J. B. Zuber, {\it Quantum Field Theory} 
(McGraw--Hill, New York, 1985)
\\
$^{3}$A. D. Linde, {\it Particle Physics and Inflationary Cosmology} 
(Harwood Academic Publishers, London, 1988)
\\
$^{4}$G. K. Savvidy, Phys. Lett. B {\bf 130}, 303 (1983); 
Phys. Lett. B {\bf 159}, 325 (1985); Nucl. Phys. B {\bf 246}, 302 (1984)
\\
$^{5}$T. Kawabe and S. Ohta, Phys. Rev. D {\bf 44}, 1274 (1991); 
Phys. Lett. B {\bf 334}, 127 (1994); 
T. Kawabe, Phys. Lett. B {\bf 343}, 254 (1995)
\\
$^{6}$M.S. Sriram, C. Mukku, S. Lakshmibala and 
B.A. Bambah: Phys. Rev. D {\bf 49}, 4246 (1994); 
J. Segar and M.S. Sriram, Phys. Rev. {\bf D 53}, 3976 (1996)
\\
$^{7}$M.A. Halasz and 
J.J.M. Verbaarschot, Phys. Rev. Lett. {\bf 74}, 3920 (1995)
\\
$^{8}$S.G. Matinyan and B. Muller, 
Phys. Rev. Lett. {\bf 78}, 2515 (1997)
\\
$^{9}$A. H. Nayfeh and B. Balachandran, 
{\it Applied Nonlinear Dynamics} (J. Wiley, New York, 1995)
\\
$^{10}$J. Guckenheimer and P. Holmes, {\it Nonlinear Oscillations, 
Dynamical Systems, and Bifurcations of Vector Fields} 
(Springer, New York, 1983)
\\
$^{11}$D. W. Jordan and P. Smith, {\it Nonlinear Ordinary Differential 
Equations} (Oxford Univ. Press, Oxford, 1987)
\\
$^{12}$L. Salasnich, Mod. Phys. Lett. A {\bf 10}, 3119 (1995)
\\
$^{13}$L. Salasnich, Nuovo Cim. B {\bf 112}, 873 (1997)
\\
$^{14}$L. Salasnich, Phys. Rev. D. {\bf 52}, 6189 (1995)
\\
$^{15}$L. Salasnich, in 
{\it Perspectives on Theoretical Nuclear Physics}, vol. {\bf 6}, 
pp. 261--268, 
Ed. A. Fabrocini {\it et al.} (Edizioni ETS, Pisa, 1996)
\\
$^{16}$L. Salasnich, Mod. Phys. Lett. A {\bf 12}, 1473 (1997)
\\
$^{17}$L. Salasnich, Phys. Atom. Nucl. {\bf 61}, 1878 (1998) 
\\
$^{18}$M.C. Gutzwiller, {\it Chaos in Classical and Quantum Mechanics} 
(Springer, Berlin, 1990)
\\
$^{19}$G. Casati and B.V. Chirikov, {\it Quantum Chaos} 
(Cambridge University Press, Cambridge, 1995)
\\
$^{20}$M. Toda, Phys. Lett. A {\bf 48}, 335 (1974)
\\
$^{21}$G. Benettin, R. Brambilla and 
L. Galgani, Physica A {\bf 87}, 381 (1977)
\\
$^{22}$I. Bendixson, Acta Math. {\bf 24}, 1 (1901)
\\
$^{23}$A. Lienard, Rev. Gen. Electr. {\bf 23}, 901 (1928)

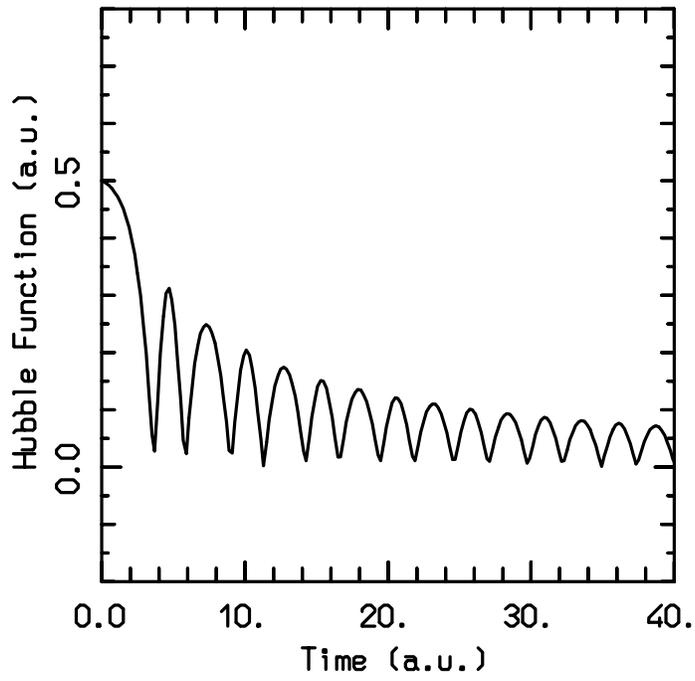
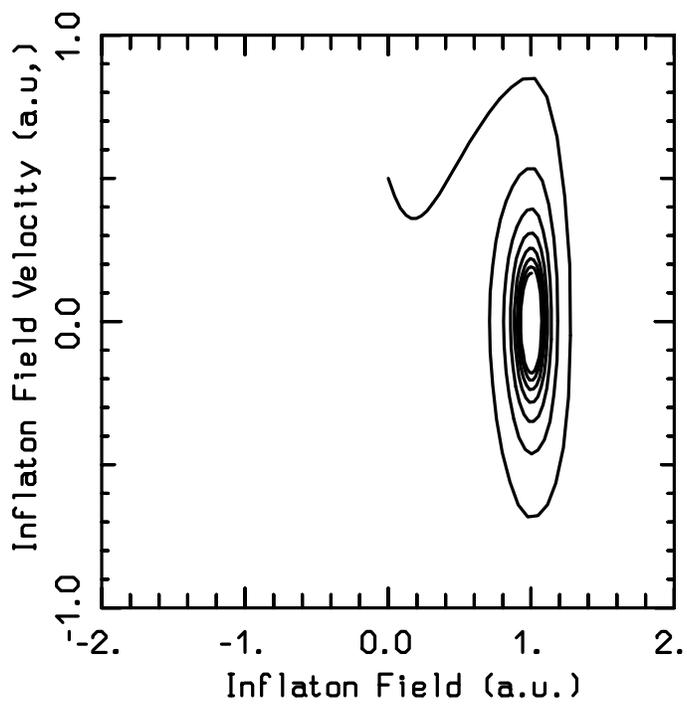
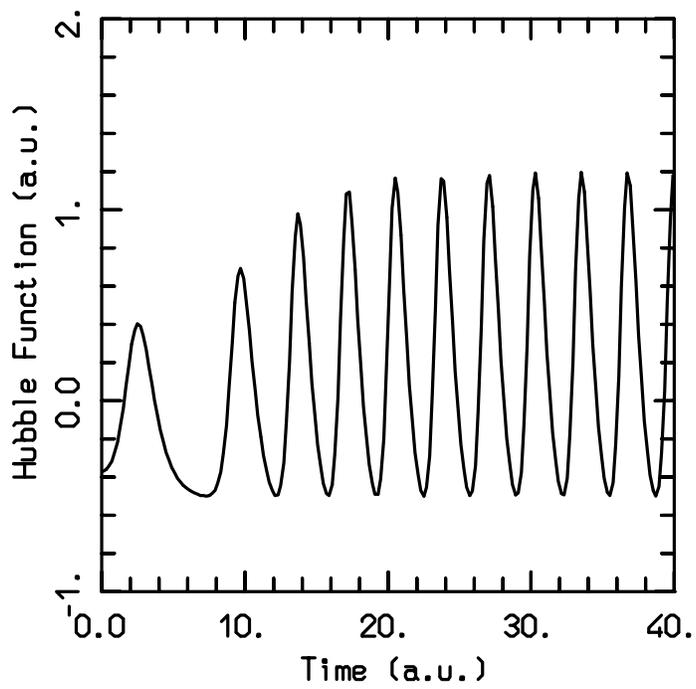
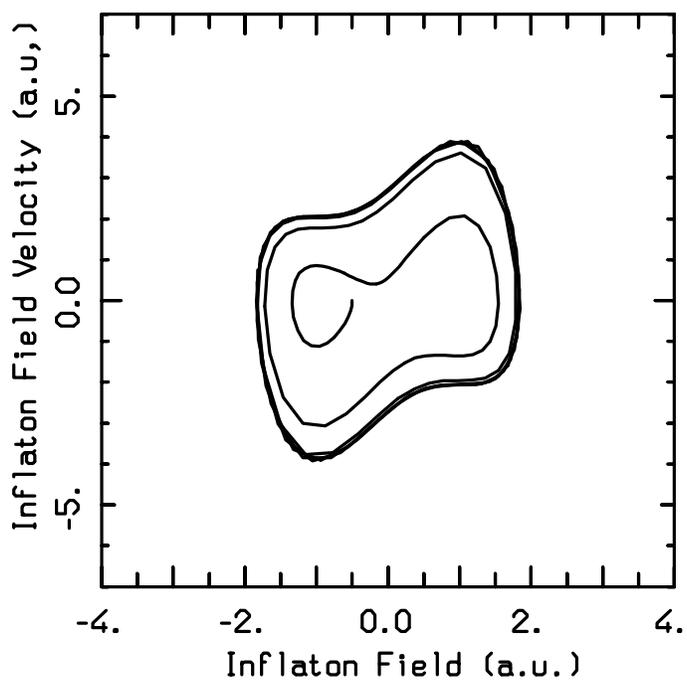
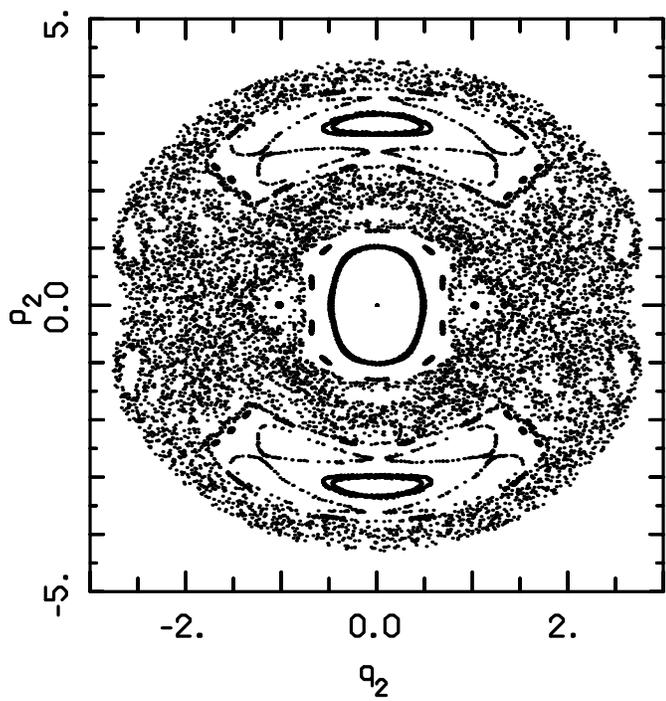
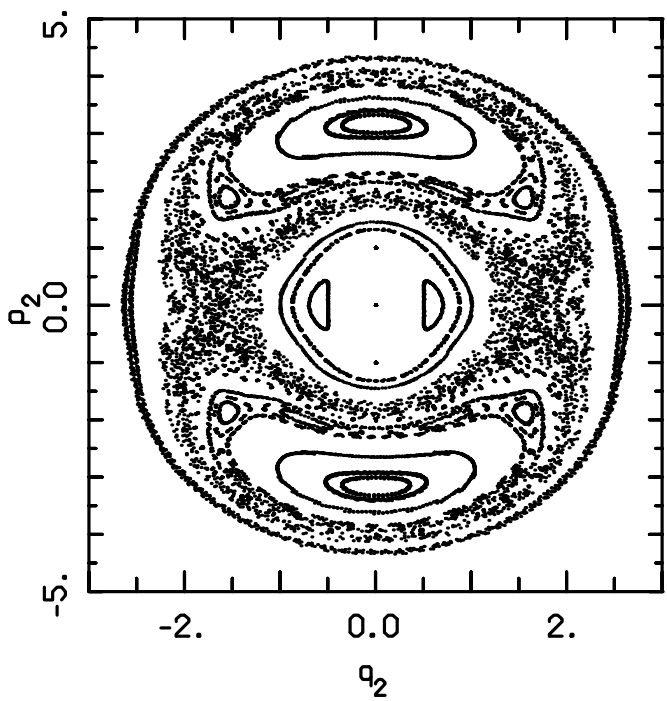
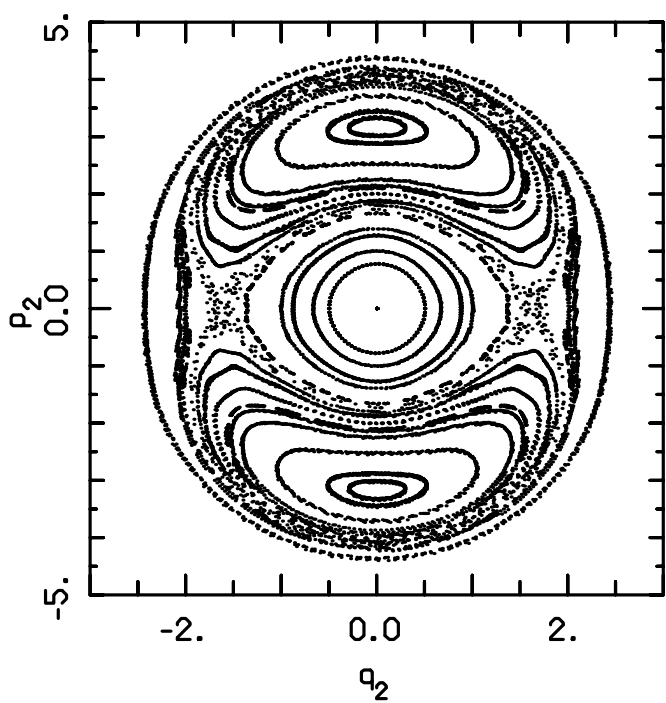
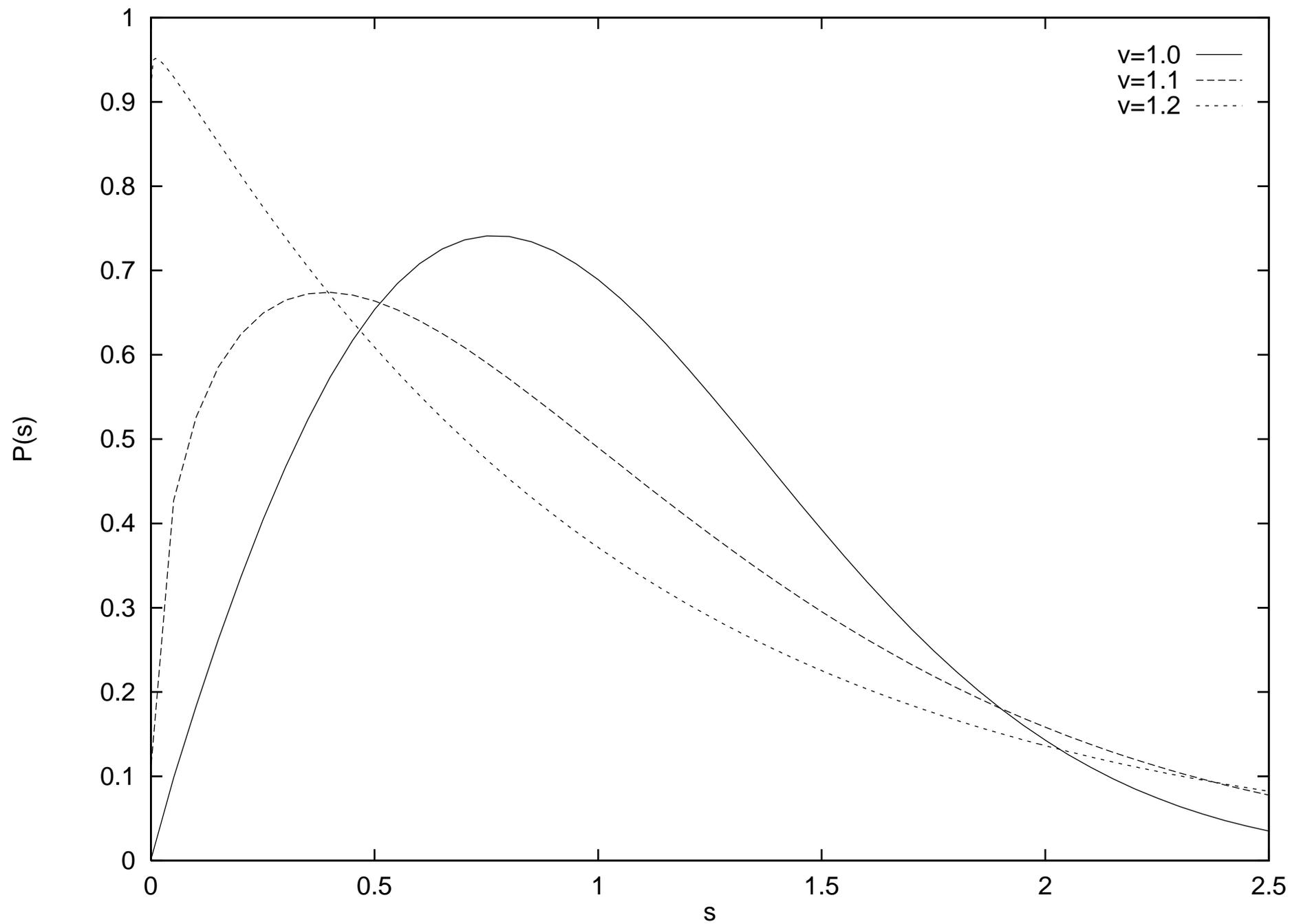
\end{document}